\pdfoutput=1
\documentclass[11pt]{article}

\usepackage[]{support/acl}
\usepackage{times}
\usepackage{booktabs}
\usepackage{latexsym}
\usepackage[T1]{fontenc}
\usepackage[utf8]{inputenc}
\usepackage{microtype}
\usepackage{graphicx}
\usepackage{todonotes}
\usepackage{url}
\usepackage{wrapfig}
\usepackage{overpic}
\usepackage{amsmath}
\usepackage{caption}
\usepackage{subcaption}
\usepackage{array}

\usepackage{color}

\newcommand\myparskip{\vspace{0.25cm}}

\title{From Numbers to Words: \\Multi-Modal Bankruptcy Prediction Using the ECL Dataset}

\author{Henri Arno$^{1*}$\ \and Klaas Mulier$^1$ \and Joke Baeck$^1$ \and Thomas Demeester$^2$ \\
  $^1$Ghent University\\
  $^2$Ghent University - imec\\
  \texttt{Henri.Arno@UGent.be} \\}

\begin{document}
\maketitle
\begin{abstract}
In this paper, we present ECL, a novel multi-modal dataset containing the textual and numerical data from corporate 10K filings and associated binary bankruptcy labels. Furthermore, we develop and critically evaluate several classical and neural bankruptcy prediction models using this dataset. Our findings suggest that the information contained in each data modality is complementary for bankruptcy prediction. We also see that the binary bankruptcy prediction target does not enable our models to distinguish next year bankruptcy from an unhealthy financial situation resulting in bankruptcy in later years.
Finally, we explore the use of LLMs in the context of our task. We show how GPT-based models can be used to extract meaningful summaries from the textual data but zero-shot bankruptcy prediction results are poor. All resources required to access and update the dataset or replicate our experiments are available on 
\url{github.com/henriarnoUG/ECL}. 
\end{abstract}

\section{Introduction}
Bankruptcy has far-reaching consequences that extend beyond the business owners, affecting various stakeholders such as employees, suppliers and creditors. On an economy-wide scale, bankruptcy risk plays a structural role in propagating recession \cite{bernanke1981bankruptcy}. Predicting the occurrence and timing of this corporate event precisely is challenging, due to the external factors and complex financial dynamics at play. Yet, certain warning signals, such as decreasing revenues and rising debt, can serve as an indication of imminent bankruptcy. Therefore, several researchers have directed their efforts towards the development of sound bankruptcy prediction models in the past decades \cite{beaver1966financial, ohlson1980financial}. Increasingly advanced prediction models \cite{odom1990neural, KIM20103373}, combined with well-chosen, informative features \cite{mai2019deep}, have led to increased predictive performance in the field.

\myparskip
\noindent
In this paper, we contribute to the literature in two ways. First, we present ECL, a new dataset that contains the textual and numerical data from corporate 10K filings (cf. Section~\ref{sec:data}) and associated binary bankruptcy labels. It is a unique compilation of three existing data sources: the \underline{E}DGAR-corpus \cite{loukas-etal-2021-edgar}, \underline{C}ompuStat\footnote{URLs accessed 2023-10-05: \\ \url{https://www.marketplace.spglobal.com/en/datasets/compustat-financials-(8)} and \\ \url{https://lopucki.law.ufl.edu}} and the \underline{L}oPucki Bankruptcy Research Database.\footnotemark[1] Second, we present baseline bankruptcy prediction models on each data modality, as well as on the combination, and critically evaluate their performance. Based on our findings, we identify and formulate interesting avenues for future research. 


\myparskip
\noindent
In recent work \cite{arno-etal-2022-next}, we argue that contributions in the field of bankruptcy prediction are difficult to compare since (1) the considered evaluation scenarios vary strongly (which is related to the temporal nature of the data and the class imbalance) (2) there is no consensus on key evaluation metrics and (3) there is a lack of benchmark datasets. We introduced a carefully designed evaluation strategy, applied to a text-only benchmark dataset. In this paper, we adopt this evaluation setup, report the suggested evaluation metrics for our baseline models on ECL, and share our code and dataset to encourage reproducible future research (see \url{github.com/henriarnoUG/ECL}). 

\myparskip
\noindent
Our findings suggest that the textual and numerical content from a 10K contain complementary information for bankruptcy prediction. In some cases, the management of a company explicitly state that they consider filing for bankruptcy, making the prediction task based on text trivial. If this is not mentioned, the accounting figures are more informative for bankruptcy prediction. Furthermore, the results show that our models trained on binary labels cannot distinguish 10K records filed in the year before bankruptcy from those records filed by financially unhealthy companies that did not file for bankruptcy just yet. Based on this finding we argue that modelling the financial health of a company with a more gradual label is an interesting direction for future research. Finally, we explore the potential of LLMs in the context of our task. We show that GPT-generated summaries from the text contained in the 10K filings are useful for bankruptcy prediction. Despite this promising result, we find that the zero-shot prediction results of GPT-3.5 are significantly worse than the results of a simple keyword-based TF-IDF model. \\

\noindent
The structure of this paper is as follows. In Section~\ref{sec:data} we present our dataset and discuss the prediction task. Section~\ref{sec:models} contains an overview of our experimental setup. The results, along with an in-depth qualitative analysis, are presented in Section~\ref{sec:results}. The potential of LLMs for our task is explored in Section~\ref{sec:gpt} while Section~\ref{sec:concl} concludes.

\section{The ECL Dataset} 
\label{sec:data} 

Large companies operating in the U.S. are required to submit a variety of filings with the Securities and Exchange Commission (SEC) throughout the year. Potential investors and other stakeholders use these filings to gain insight into the financial performance, business operations, risks and other aspects of the company of interest. Notably, the most widely consulted SEC filing is the Form 10K, which is reported annually and contains detailed information on a company's past fiscal year. A 10K filing is composed of 15 different items including a description of the business (item 1), the management discussion and analysis (item 7) and a section on executive compensation (item 11), among others. Item 8 of a Form 10K contains the consolidated financial statements such as the balance sheet, the income statement and the cashflow statement. We carefully compiled the EDGAR-CompuStat-LoPucki dataset, further referred to as ECL, containing data in two modalities (i.e., textual and numerical) 
from such 10K records that companies filed with the SEC in the past. 
We present the dataset in the context of our current work on bankruptcy prediction, but are convinced that the multi-modal dataset has other possible uses, in terms of analysis or predictive modelling of a companies' financial and business situation.

\subsection{Data Sources}
\label{sec:sources}
Most SEC filings, including the Form 10K, are publicly available through the Electronic Data Gathering, Analysis and Retrieval (EDGAR) website as a text file or as an XBRL file (an HTML based document type). The same 10K data 
can be accessed 
through a variety of other sources. ECL is a unique compilation of three existing data 
sources: the textual data is collected from (1) the \underline{E}DGAR-corpus \cite{loukas-etal-2021-edgar}, the numerical financial data is gathered from (2) \underline{C}ompuStat\footnote{In order to use ECL, access to CompuStat is required. For details, we refer to our GitHub repository.} while (3) the \underline{L}oPucki BRD provides the labels for the bankruptcy prediction task.

\subsection{Dataset Construction and Labelling}
\label{sec:constr}

Some firms are required to file a 10K every year, such as companies whose stock is traded on a U.S. stock exchange, while others voluntarily submit 10K filings. Using the EDGAR-crawler tool,\footnote{Available at: \\ \url{github.com/nlpaueb/edgar-crawler}} we have collected the textual data (and corresponding metadata) from 
all 10K filings on the EDGAR website from 1993\footnote{This is the starting point of the EDGAR-corpus as well.} onwards. \\

\noindent
Next, we add the structured, financial information, reported in item 8 of a 10K, to the dataset by linking the collected 10K records from the previous step to CompuStat records. We use the CompuStat Fundamentals North-America table and filter out the records that originate from sources other than the Form 10K (some records are collected from the prospectus, the annual letter to the shareholders, Form 20-F, ... etc.). We merge the collected 10K records and the filtered CompuStat records based on two conditions. First, matching records must have the same company name or company identifier (the Central Index Key) and second, the fiscal year end (the date) of the records must lie within 7 days of each other.\footnote{Our analysis revealed that, due to data quality issues, the fiscal year end can be a couple of days off in CompuStat.} Remaining 10K records or CompuStat records left unmatched are discarded.\\

\noindent
Finally, we assign the labels for the next year bankruptcy prediction task following our proposed labelling strategy \cite{arno-etal-2022-next}. We collect the bankruptcy data from the LoPucki Bankruptcy Research Database (BRD). This dataset contains the exact date on which companies filed for bankruptcy under chapter 7 or chapter 11 of the U.S. bankruptcy code. Only firms that (1) submit 10K filings with the SEC and (2) have a total asset value exceeding 100,000,000, measured in 1980 dollars, qualify for inclusion in the LoPucki BRD. Data is available for all bankruptcies between 1979 and the end of 2022. Before we assign labels, we tag the 10K filings in ECL based on these criteria. The total asset value reported in the 10K must exceed the (inflation-corrected) threshold and the filing date must lie within the correct time frame. \\ 

\begin{figure}[]
    \centering
    \vspace{0cm}
    \begin{overpic}[width=0.4\textwidth]{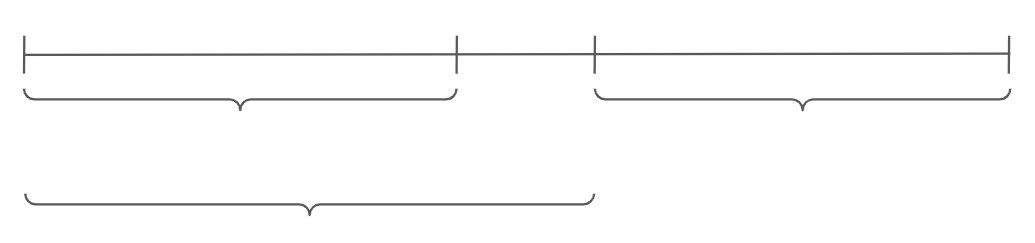}
    \put(0,25){t\textsubscript{PR - 1 year}}
    \put(54,25){t\textsubscript{FD}}%
    \put(40,25){t\textsubscript{PR}}%
    \put(84,25){t\textsubscript{FD + 1 year}}%
    \put(72,7){T\textsubscript{Pred}}%
    \put(20,7){T\textsubscript{PR}}%
    \put(25,-5){T\textsubscript{Ind}}%
    \end{overpic}
    \vspace{0.3cm}
    \caption{\label{fig:label} The labelling strategy for the 10K records in our dataset for the next year bankruptcy prediction task.}
\end{figure}
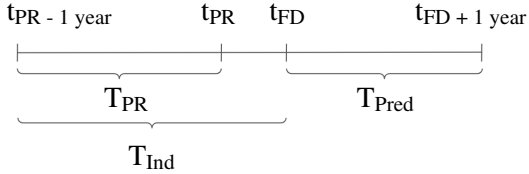

\noindent
Each qualified 10K record in the dataset is assigned a binary label. A 10K filing covers a fiscal year (T\textsubscript{PR}), which concludes on the fiscal year end (t\textsubscript{PR}), and is released to the public on the filing date (t\textsubscript{FD}), after the filing period. The bankruptcy label is true if the company filed for bankruptcy in the year following the filing date (i.e. during T\textsubscript{Pred}) and false otherwise. The task is to predict whether a company will file for bankruptcy in the next year, given the multi-modal data contained in the 10K filing (covering the period T\textsubscript{Ind}). This labelling strategy is graphically depicted in Figure~\ref{fig:label}.


\subsection{ECL Statistics}
\label{sec:stats}
ECL consists of 170,139 Form 10K filings for which numerical and textual data is available. From the the 277,940 collected 10K filings and the 241,825 filtered CompuStat records, 107,801 and 71,686 records remain unmatched, respectively. The vast majority (over 56\%) of these unmatched records come from companies with a standard industrial classification (SIC) code in the finance, insurance or real estate division. Some examples include investment offices (24.2\% of the unmatched records) or companies issuing asset-backed securities (12.5\% of the unmatched records). The distribution of the 10K records in our dataset over the different industries (SIC divisions) can be found in Table~\ref{tbl:sectors} in the Appendix. \\

\vspace{-0.23cm}
\noindent
The 10K filings in our dataset come from 18,582 unique companies for which 
we have 9.16 years of data on average. These companies are relatively large with an average total asset value of 1.39 billion dollars\footnote{After removal of outliers exceeding the 95\% quantile.} and are well distributed across the United States as can be seen in Figure~\ref{fig:map} in the Appendix. The state where most companies have their headquarters  is California, followed by Texas, New York and Florida. The 10K filings in the dataset are relatively long, consisting on average of 29,247 words. The longest items in the 10K filings are item 7: the management discussion and analysis or the MD\&A (6,810 words on average), item 1: the business description (6,123 words on average) and item 15: the exhibits (4,799 words on average).

\subsection{Data Splits for Bankruptcy Prediction}
From the  170,139 records in our dataset, 84,652 qualify for inclusion in the LoPucki BRD (cf. Section~\ref{sec:constr}) and were assigned a binary label. Among these 10K records, 662 were filed in the year preceding bankruptcy (i.e. the positives) while 83,990 were not. This implies a strong class imbalance with about 1 positive sample for every 127 negative samples. The labelled 10K records filed prior to 2012 are used to train the models while the records filed between 2012 and 2015 are assigned to the validation set, which is used for hyperparameter optimisation and model selection. The remaining 10K records, filed after 2015, make up the test set and are used to evaluate the final models (which are retrained on all 10K's in the train and validation set). The train, validation and test sets consists of 54,039; 12,324 and 18,289 filings respectively with 481, 59 and 122 positive cases each. For an overview of the splits, see Table~\ref{tbl:overviewECL} in the Appendix.

\section{Experimental Setup}
\label{sec:models}

As discussed above, a 10K record consists of various items and contains different data modalities. First, we separately explore the predictive value of (1) the numerical financial data of the 10K's and (2) the text in the reports, specifically from item 7: the management discussion and analysis. Afterwards, we build a predictive model that uses both data types jointly. In this section we cover the design of the models and briefly discuss the training details.

\subsection{Numerical Models}
\label{sec:num-models}
The consolidated financial statements are reported in item 8 of a 10K and contain a large number of financial figures. For our prediction models, we employ the most informative accounting figures in line with previous work \cite{mai2019deep}.\footnote{We discard the market-based predictors (e.g. stock market returns) used by \citeauthor{mai2019deep} (\citeyear{mai2019deep}) and only use those features that can be computed from the 10K.} In Table~\ref{tbl:description} in the Appendix we give an overview of the variables that serve as an input for our classifiers. As a baseline, we train a logistic regression classifier with an L2-regularisation penalty. Furthermore, we include a multi-layer perceptron and an XGBoost classifier \cite{chen2016xgboost} as more advanced alternatives. For the logistic regression model, we only tune the regularisation strength. The dimensions of the hidden layer(s), the learning rate and the regularisation strength are the hyperparameters of the MLP. For the XGBoost model, we optimise the number of trees, the shrinkage factor, the proportion of the data to sample at each split and the maximum depth of the trees.\\


\vspace{-0.2cm}
\noindent
Due to the infrequent occurrence of bankruptcy, our dataset is heavily imbalanced. As we want our models to be able to discriminate between bankrupt and non-bankrupt firms, we need a strategy to deal with the small number of positive samples (i.e. the 10K records filed in the year preceding bankruptcy). Therefore, we randomly oversample the minority instances in our training data and treat the ratio of positive over negative samples as a hyperparameter as well. Furthermore, we impute missing values 
(except for the XGBoost model that can handle missing data), centre the variables around the mean and scale them to unit variance. For each model, we set the values of the hyperparameters that maximise the area under the receiver operating curve (ROC-AUC)\footnote{We do not report the results of the models when tuned on average precision (AP) instead of ROC-AUC as there was no substantial difference.} (cf. \citeauthor{mai2019deep} (\citeyear{mai2019deep}) and \citeauthor{arno-etal-2022-next} (\citeyear{arno-etal-2022-next})) (see Section~\ref{sec:pm} for more details on this performance metric). 

\subsection{Textual Models}
\label{sec:text-models}
A Form 10K is an extensive document. On average, a filing in our dataset has 29,247 words. However, much of this content is not relevant for our prediction task (such as the description of the business or the exhibits). The most informative part of the 10K can be found in item 7: the management discussion and analysis. In this section, the management of the company gives its view on the past fiscal year, discussing the risks that the company faced, special circumstances that had an effect on the company and many other interesting aspects that may have had an impact on the results. Consistent with prior literature \cite{cecchini2010making, mayew2015md, mai2019deep, arno-etal-2022-next}, we use the text from this part of the 10K in our prediction models. \\

\noindent
First, we train an L1-regularised logistic regression classifier that uses \textit{Term Frequency - Inverse Document Frequency} (TF-IDF) features as input. This keyword-based document representation technique has achieved good performance in information retrieval and document classification tasks \cite{manning} and serves as our baseline. The regularisation strength and the size of the n-grams are treated as hyperparameters. Second, we finetune a pretrained RoBERTa-large model on our classification task \cite{zhuang-etal-2021-robustly}. We only pass the first 512 tokens to the model, which corresponds to its maximum sequence length. In the first epoch, we train only the classification head 
and freeze the parameters of the encoder. For the second and last epoch, we adjust the learning rate downwards and train the entire model. We use a batch size of 320 instances.
In order to handle the class imbalance, we weigh the samples inversely proportional to the class frequencies during training of each textual model. 

\begin{table*}[t]
\vspace{-0.8cm}
\centering
\begin{tabular}{@{}l|ccc|cc|c@{}}
\toprule
\textbf{Data Modality} & \multicolumn{3}{c|}{Numeric}             & \multicolumn{2}{c|}{Textual} & Combined         \\ \midrule
\textbf{Model}        & LogReg & MLP            & \textbf{XGBoost}        & \textbf{TF-IDF}           & RoBERTa   & \textbf{XGBoost + TF-IDF} \\ \midrule
\textbf{ROC-AUC}       & 0.915  & 0.925          & \textbf{0.936} & \textbf{0.886}   & 0.796     & \textbf{0.948}   \\
\textbf{AP}            & 0.115  & \textbf{0.162} & 0.156          & \textbf{0.239}   & 0.067     & \textbf{0.264}   \\
\textbf{Recall@100}    & 0.148  & \textbf{0.197} & 0.189          & \textbf{0.287}   & 0.107     & \textbf{0.287}   \\
\textbf{CAP ratio}     & 0.830  & 0.851          & \textbf{0.873} & \textbf{0.771}   & 0.591     & \textbf{0.896}   \\ \bottomrule
\end{tabular}
\caption{\label{tbl:res} The results of the numerical, textual and combined models, tuned on ROC-AUC, evaluated on the test set. For each data modality, the best result is shown in bold.}
\end{table*}

\subsection{Combined Numerical and Textual Model}
\label{sec:Combined}
\vspace{-0.01cm}
To leverage the combined predictive power of both the numerical and textual data, we employ an ensemble model. By combining the outputs of the best uni-modal classifiers, we aim to achieve the best overall predictive performance for the 
bankruptcy prediction task. In our ensemble approach, we retrain the best uni-modal models on the train set and have them score the instances in the validation set. Similarly to stacked generalisation \cite{wolpert1992stacked}, the normalised scores are then used to train a meta-classifier that makes the final prediction. Finally, we can use the base classifiers to make uni-modal predictions on the test set, which are used by the meta-classifier to generate a final prediction, taking both data modalities into account.

\section{Bankruptcy Predictions Result and Analysis}
\label{sec:results}

In this section we motivate the choice of performance metrics, report the results of the models on the test set, trained with optimal hyperparameter values, and discuss our most interesting findings.



\subsection{Performance Evaluation}
\label{sec:pm}
\vspace{-0.1cm}
We report the area under the receiver operating curve (ROC-AUC), the average precision (AP), the cumulative accuracy profile ratio (CAP ratio) and the recall@100 for each classifier. The \textbf{ROC-AUC} summarises the ROC curve, which shows the true positive rate and false positive rate at each classification threshold. The ROC-AUC can be interpreted as the probability that a randomly chosen positive instance (i.e. a 10K record filed in the year before bankruptcy) is scored higher than a randomly chosen negative one by the classifier \cite{fernandez2018learning}. The \textbf{AP} is a metric that summarises the precision-recall (PR) curve and reflects the performance of the model on the minority class. The PR curve graphically depicts the trade-off between precision and recall at each classification threshold. The average precision metric (AP) is particularly valuable when dealing with highly skewed data distributions where the ROC-AUC can be overly optimistic \cite{davis2006relationship}. The \textbf{recall@100} gives the proportion of positives, retained in the 100 highest ranked instances by the classifier, out of all positives. In our application, it reflects the ability of a model to detect 10K records filed in the year preceding bankruptcy given a fixed budget (i.e., when only 100 filings can be retrieved). Finally, we report the \textbf{CAP ratio}, a metric that summarises the cumulative accuracy profile curve \cite{mai2019deep}. This curve shows the recall at varying percentages of observations when sorted according to the classifiers' scores. 
Furthermore, we also show the PR, ROC and CAP curves for the best numerical, textual and combined models.

\subsection{Bankruptcy Classification Performance}


From the results in Table~\ref{tbl:res}, we conclude that the MLP and XGBoost classifiers achieve the best performance among the models trained on numerical predictors. The MLP classifier has the best average precision (AP) and recall@100 while the XGBoost model has the highest ROC-AUC and CAP ratio. Furthermore, within the class of models trained solely on text, the keyword-based TF-IDF model attains the best results on all performance metrics. It is worth noting that this is the only model capable of processing the entire documents, unlike RoBERTa, which has a maximum sequence length. As expected, the best results overall are attained by the ensemble model, which leverages both data modalities and combines the predictions from the XGBoost and TF-IDF classifiers. \\ 

\begin{figure*}[]
\footnotesize
\vspace{-1cm}
\centering
\begin{subfigure}{.5\textwidth}
  \centering
  \includegraphics[width=\textwidth]{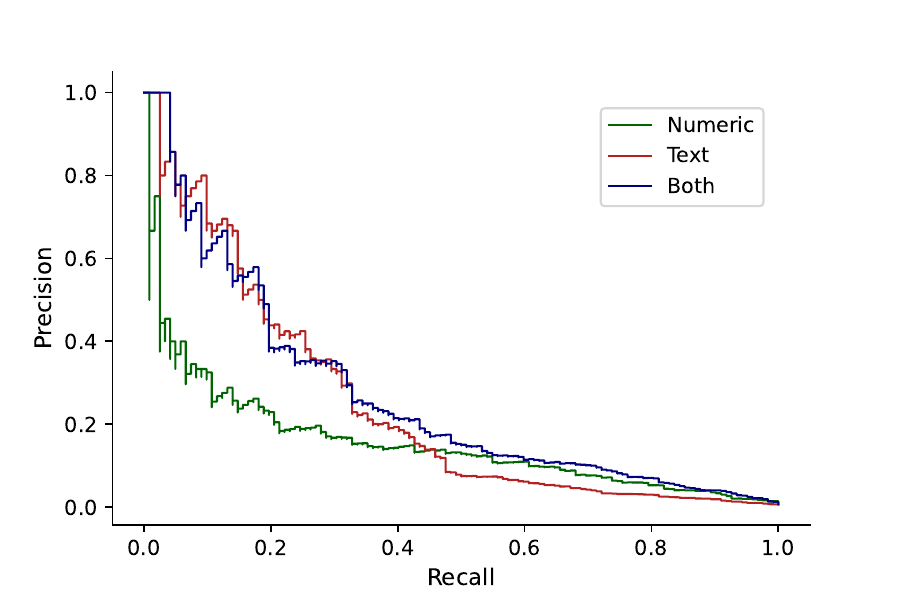}
  \caption{Precision-Recall Curve (PR curve)}
  \label{fig:PR}
\end{subfigure}%
\begin{subfigure}{.5\textwidth}
  \centering
  \includegraphics[width=\textwidth]{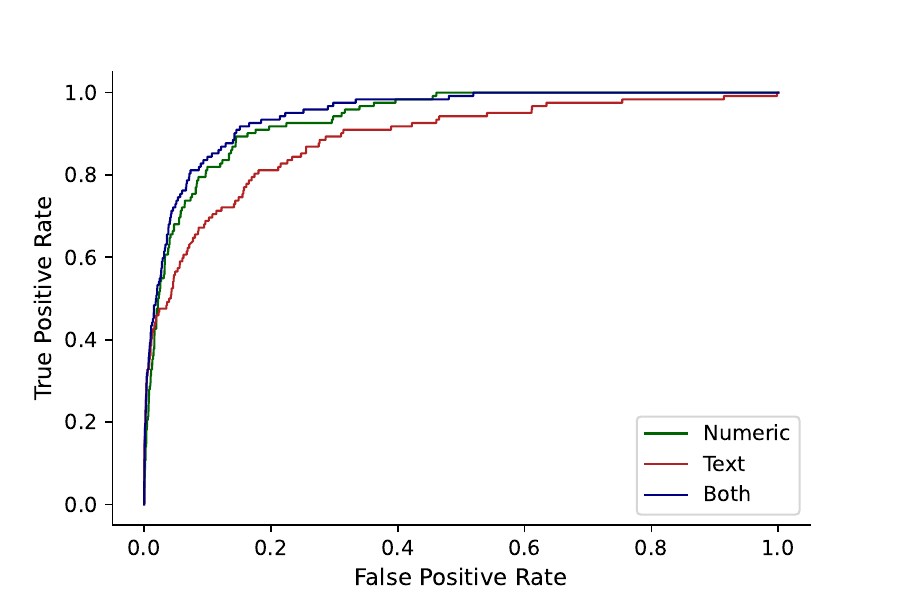}
  \caption{Receiver Operating Curve (ROC curve)}
  \label{fig:ROC}
\end{subfigure}
\caption{Precision-Recall Curve and Receiver Operating Curve for the best textual (TF-IDF), numerical (XGBoost) and combined (ensemble) models evaluated on the test set.}
\label{fig:Perf}
\end{figure*}

\noindent
The PR curve shown in Figure~\ref{fig:PR} provides additional insights into the performance of the best numerical (XGBoost), textual (TF-IDF) and combined (ensemble) model. We can see that, at low classification thresholds (i.e., when bankruptcy predictions for 10K filings are infrequent), the textual and combined models are comparable while the numerical model lags behind. As the threshold increases and more instances are classified as bankrupt, the precision of the textual models drops quickly and the performance of the numerical model becomes on par with the combined model. The ROC and CAP curves, shown in Figure~\ref{fig:ROC} and Figure~\ref{fig:CAP} in the Appendix, support our previous results and consistently display the best performance for the combined model. The numerical model follows closely while the textual model comes in last.

\begin{table}[]

\begin{tabular}{@{}p{7.5cm}}
\toprule
\textbf{Snippet from item 7: management discussion and analysis} \\ \midrule
\textit{"...we also may conclude that it is necessary to initiate proceedings under Chapter 11 of the United States Bankruptcy Code..."}                  \\ \midrule
\textit{"...it may be necessary for us to seek protection from creditors under Chapter 11 of the U.S. Bankruptcy Code..."}                                     \\ \midrule
\textit{"...it may be necessary to seek a private restructuring or protection from creditors under Chapter 11 of the United States Bankruptcy Code..."} \\ \bottomrule
\end{tabular}
\caption{\label{tbl:examples} Snippets from the MD\&A of the top ranked instances in our test set by the TF-IDF model.}
\end{table}

\newpage
\subsection{Qualitative Analysis of the Results}
\paragraph{Complementary Information in the Textual and Numerical Data:} A first interesting result is that the TF-IDF model has a better recall@100 but a worse CAP ratio than any classifier trained on numerical data. This suggests that some specific 10K filings are more easily classified when using text from the MD\&A as input compared to using the accounting figures. A qualitative inspection of the 10K filings that were ranked highest by the TF-IDF model, revealed that the management of the company sometimes explicitly states that they consider filing for bankruptcy in the coming year. The snippets in table \ref{tbl:examples} show this behaviour for the 3 highest ranked 10K filings, by the TF-IDF model, in our test set. This 
information cannot be directly quantified in any of the parameters of the numerical 
models, and in that respect, we can conclude that the information contained in both data modalities has some complementary value.

\noindent
As an alternative way of investigating the importance of the textual vs.~numerical data, we analyse the combined numerical and textual model introduced in Section~\ref{sec:Combined}. In particular, we
trained a logistic regression classifier with three parameters on the normalised scores from the XGBoost and the TF-IDF classifiers. The weights ($\beta_1$ and $\beta_2$ in the equation below) represent the relative importance of the information contained in each data modality for the bankruptcy predictions by our ensemble, the best performing model overall.
\begin{gather*}
    P(\text{\textit{Next year bankruptcy}}) \\
    = \sigma(\beta_0 + \beta_1 Score_{XGBoost} + \beta_2 Score_{TF-IDF}) 
\end{gather*}

\noindent
The $\beta_1$ and $\beta_2$ coefficients are 1.30 and 0.321 respectively, indicating that the numerical data is most informative for the task at hand. This finding supports our previous result, that the textual data is mainly useful for the classification of those few 10K filings where the consideration for bankruptcy is clearly stated. Remaining 10K records are better classified using the accounting figures. 

\noindent
\paragraph{Ranking Performance of the Models:} 
As seen from Table~\ref{tbl:res}, our models 
attain high values on ROC-AUC and the  CAP ratio.  
This can be directly linked to the imbalanced nature of the problem, since only a very small 
portion of 10K records are filed in the year before bankruptcy. For the ROC-AUC metric, 
this means that, for a sufficiently large $k$, the top $k$ highest ranked 
instances, 
by any considered model, would contain a large fraction of all 
records filed in the year before bankruptcy (leading to a high recall or true positive rate), 
and a small fraction of all records not filed in the year before bankruptcy (i.e., a low false positive rate or FPR). 
This is illustrated by the ROC curves in Figure~\ref{fig:ROC}. 
In that same set of $k$ highest scoring instances, the number of records filed in the year before bankruptcy is however relatively small compared to the number of records not filed in the year before bankruptcy, leading in turn to a low precision at $k$, and by extension, a low AP metric. This is illustrated in the precision-recall curve in Figure~\ref{fig:PR}. When considering the ensemble model and $k=100$, 
35 positives and 65 negatives are retained from the total of 122 positives and 18,167 negatives in the test set (corresponding to a recall@100 of 28.7\%, 
and a FPR of 0.3\%).\\

\vspace{-0.2cm}
\noindent
This result can be further nuanced in light of casting the problem as a binary classification task, which is common in bankruptcy prediction literature. Our hypothesis was, that from these 65 false positives (again considering the top 100 highest ranked 10K records by our best model), some 10K's were filed by companies worthy of further investigation due to their unhealthy financial situation, although they just did not quite file for bankruptcy in the following year yet. 
Indeed, the 65 false positive 10K records were filed by 53 different companies, of which an additional 12 turned out to have filed for bankruptcy by 2023. Modelling financial health 
with a more gradual label can therefore be expected to lead to a higher consistency, although the concept of financial health itself is less straightforward to quantify unambiguously. We consider this a highly useful direction for future research.

\begin{table*}[]
\vspace{-0.8cm}
\centering
\begin{tabular}{l|cc|ccc}
\hline
\textbf{Data Modality} & \multicolumn{2}{c|}{Textual: GPT summaries} & \multicolumn{3}{c}{Textual: Full MD\&A} \\ \hline
\textbf{Model}      & TF-IDF & \textbf{RoBERTa} & \textbf{TF-IDF} & RoBERTa & GPT-3.5 (zero-shot) \\ \hline
\textbf{ROC-AUC}    & 0.893      & \textbf{0.902}       & \textbf{0.912}      & 0.592       & 0.667 ($\pm$0.022)       \\
\textbf{AP}         & 0.089      & \textbf{0.202}       & \textbf{0.294}      & 0.021       & 0.019 ($\pm$0.001)      \\
\textbf{Recall@100} & \textbf{0.600}      & 0.500       & \textbf{0.700}    & 0.200       & 0.148 ($\pm$0.050)       \\
\textbf{CAP ratio}  & 0.791      & \textbf{0.804}       & \textbf{0.824}      & 0.184       & 0.335 ($\pm$0.044)       \\ \hline
\end{tabular}
\caption{\label{tbl:gptres} The results on the randomly sampled test (of 1000 instances) of the textual models trained on the extracted summaries, the original textual models and GPT-3.5 (zero-shot). Due to the different sizes of this sampled test set and the original test set, the values in this table and Table~\ref{tbl:res} are not directly comparable.}
\end{table*}

\noindent

\section{The Potential of LLMs for Text-Based Bankruptcy Prediction}
\label{sec:gpt}

\paragraph{GPT Prompting Strategy}
Recently, large language models (LLMs) have shown to be tremendously successful on a variety of tasks, including financial text classification \cite{loukas2023breaking} and zero-shot text summarisation \cite{goyal2022news}. In this section, we explore how such models can be used in the context of bankruptcy prediction. More specifically, we will use GPT-3.5 Turbo \cite{ouyang2022training} with a context window of 16,000 tokens to (1) summarise the text from the MD\&A section of the 10K filings into a single paragraph and (2) for zero-shot bankruptcy prediction. 
Due to the associated costs, we do not perform the GPT-based experiments on the entire dataset. Instead, we sample a balanced train set and a random test set from ECL of 1000 instances each. \\

\noindent
Using a single prompt (shown in Figure~\ref{fig:prompt} in the Appendix), we ask the model to summarise the MD\&A, with a particular focus on the elements that are indicative for the financial health of the company, and to assign a score, ranging from 1 to 10, that indicates how likely it is that the company will file for bankruptcy in the next year. The extracted summaries are then used to re-train the TF-IDF baseline and the RoBERTa model, which is now able to use a compact version of the entire document instead of only the first 512 tokens. We use the same training details as before with two exceptions. Since our sampled training set is balanced, we no longer use a weighted loss function and reduce the batch size to 16. For the zero-shot bankruptcy prediction task, we extract the scores that GPT-3.5 assigned to each document in the test set,\footnote{We were able to extract a score for over 83\% of the instances in the test set.} rank the test set accordingly and calculate the performance metrics. Since many instances are assigned the same score, we repeat this process 50 times and randomly shuffle documents with the same score in the ranked test set, to quantify the level of variation in the metrics due to the discrete nature of the GPT-assigned scores. The results of the models trained on the summaries, the original models and GPT-3.5 zero-shot scores on the sampled test of 1000 instances are reported in Table~\ref{tbl:gptres}. \\

\paragraph{Summarisation and Zero-Shot Bankruptcy Prediction Performance}
From the results in Table~\ref{tbl:gptres}, we can see that the TF-IDF model, trained on the complete text of the MD\&A, is still the best textual model overall. GPT-3.5 (zero-shot), the only other model capable of processing entire documents, does significantly worse. An inspection of the top ranked instances by the TF-IDF model, from the sampled test set, showed once again that the good performance of the model can be attributed to its ability to detect 10K filings where the management of the company states that they consider to file for bankruptcy in the next year (cf. Table~\ref{tbl:examples}). \\

\noindent
The results also show that the summaries extracted by GPT-3.5 are informative for the bankruptcy prediction task. The performance of RoBERTa increased tremendously when trained on these summaries instead of the first 512 tokens of the MD\&A. The performance of the TF-IDF model decreased slightly. This is not surprising, 
since the summaries contain less information than the complete MD\&A and might not capture the sentences where the management states that they consider filing for bankruptcy in the next year. Also, the models trained on the summaries saw only a fraction of the number of training instances compared to the models that saw all of the full-text training instances. Notice how RoBERTa achieves even better performance than the TF-IDF model when both are trained on the summaries, showcasing the strength of the model on a limited context. \\

\noindent
In conclusion, the summaries extracted by GPT contained useful information for bankruptcy prediction but the model performed poorly in the zero-shot setting. Additionally, we acknowledge that the quality of the summaries varied and that we encountered some samples where the model suffered from hallucination. In some rare cases, the MD\&A is not part of the 10K filing but it is included in another document (such as the annual letter to the shareholders) and item 7 of the 10K contains only a single sentence referencing this document. The GPT-generated summaries in these cases were a paragraph long and contained only imaginary facts. We believe that the performance of LLMs, in terms of summarisation and zero-shot bankruptcy prediction, can be further increased with some additional effort, but that lies outside the scope of this paper.

\section{Conclusion}
\label{sec:concl}
In this paper, we present  ECL, a multi-modal dataset of textual and numerical data from corporate 10K filings and associated binary bankruptcy labels. We also present several classical and neural bankruptcy prediction models and provide an in-depth qualitative analysis of the results. 
\newpage

\noindent
First, our findings highlight the complementarity of the information contained in both data modalities. In the text, the management of the company sometimes explicitly states that they consider filing for bankruptcy in the coming year, making the prediction task trivial for a keyword-based TF-IDF model. If this is not mentioned, the financial numerical features are better predictors for bankruptcy. The best results are attained when we combine the predictions of the textual and numerical models in an ensemble. \\

\noindent
Second, we argue that our models achieve acceptable prediction levels that may prove useful in actual applications such as the automated screening of companies' financial status, although there clearly is room for further research on ECL. We did observe that our models, trained on binary bankruptcy labels, cannot distinguish between 10K records filed in the year preceding bankruptcy and records filed by financially unhealthy companies that are close to bankruptcy but not within one year.
This indicates that modelling the financial health of a company using more fine-grained prediction targets, is an interesting avenue for future research as well. \\

\noindent
Finally, we study the potential of LLMs in the context of bankruptcy prediction. We observe that the zero-shot bankruptcy prediction results of the GPT-3.5 model are poor. Nonetheless, owing to the large context window of the model, we demonstrate its value by using the LLM to extract meaningful summaries of the text in the 10K's for the bankruptcy prediction task.

\section*{Acknowledgements}
This research was made possible through the financial support provided by the Research Foundation Flanders (FWO) under grant number G006421N.

\noindent

\newpage
\bibliography{support/anthology,support/custom}

\begin{thebibliography}{19}
\expandafter\ifx\csname natexlab\endcsname\relax\def\natexlab#1{#1}\fi

\bibitem[{Arno et~al.(2022)Arno, Mulier, Baeck, and Demeester}]{arno-etal-2022-next}
Henri Arno, Klaas Mulier, Joke Baeck, and Thomas Demeester. 2022.
\newblock \href {https://doi.org/10.18653/v1/2022.finnlp-1.25} {Next-year bankruptcy prediction from textual data: Benchmark and baselines}.
\newblock In \emph{Proceedings of the Fourth Workshop on Financial Technology and Natural Language Processing (FinNLP)}, pages 187--195, Abu Dhabi, United Arab Emirates (Hybrid). Association for Computational Linguistics.

\bibitem[{Beaver(1966)}]{beaver1966financial}
William~H Beaver. 1966.
\newblock Financial ratios as predictors of failure.
\newblock \emph{Journal of accounting research}, pages 71--111.

\bibitem[{Bernanke(1981)}]{bernanke1981bankruptcy}
Ben~S Bernanke. 1981.
\newblock Bankruptcy, liquidity, and recession.
\newblock \emph{The American Economic Review}, 71(2):155--159.

\bibitem[{Cecchini et~al.(2010)Cecchini, Aytug, Koehler, and Pathak}]{cecchini2010making}
Mark Cecchini, Haldun Aytug, Gary~J Koehler, and Praveen Pathak. 2010.
\newblock Making words work: Using financial text as a predictor of financial events.
\newblock \emph{Decision Support Systems}, 50(1):164--175.

\bibitem[{Chen and Guestrin(2016)}]{chen2016xgboost}
Tianqi Chen and Carlos Guestrin. 2016.
\newblock Xgboost: A scalable tree boosting system.
\newblock In \emph{Proceedings of the 22nd ACM SIGKDD international conference on knowledge discovery and data mining}, pages 785--794.

\bibitem[{Davis and Goadrich(2006)}]{davis2006relationship}
Jesse Davis and Mark Goadrich. 2006.
\newblock The relationship between precision-recall and roc curves.
\newblock In \emph{Proceedings of the 23rd international conference on Machine Learning}, pages 233--240.

\bibitem[{Fern{\'a}ndez et~al.(2018)Fern{\'a}ndez, Garc{\'\i}a, Galar, Prati, Krawczyk, and Herrera}]{fernandez2018learning}
Alberto Fern{\'a}ndez, Salvador Garc{\'\i}a, Mikel Galar, Ronaldo~C Prati, Bartosz Krawczyk, and Francisco Herrera. 2018.
\newblock \emph{Learning from imbalanced data sets}, volume~10.
\newblock Springer.

\bibitem[{Goyal et~al.(2022)Goyal, Li, and Durrett}]{goyal2022news}
Tanya Goyal, Junyi~Jessy Li, and Greg Durrett. 2022.
\newblock News summarization and evaluation in the era of gpt-3.
\newblock \emph{arXiv preprint arXiv:2209.12356}.

\bibitem[{Kim and Kang(2010)}]{KIM20103373}
Myoung-Jong Kim and Dae-Ki Kang. 2010.
\newblock \href {https://doi.org/https://doi.org/10.1016/j.eswa.2009.10.012} {Ensemble with neural networks for bankruptcy prediction}.
\newblock \emph{Expert Systems with Applications}, 37(4):3373--3379.

\bibitem[{Loukas et~al.(2021)Loukas, Fergadiotis, Androutsopoulos, and Malakasiotis}]{loukas-etal-2021-edgar}
Lefteris Loukas, Manos Fergadiotis, Ion Androutsopoulos, and Prodromos Malakasiotis. 2021.
\newblock \href {https://doi.org/10.18653/v1/2021.econlp-1.2} {{EDGAR}-{CORPUS}: Billions of tokens make the world go round}.
\newblock In \emph{Proceedings of the Third Workshop on Economics and Natural Language Processing}, pages 13--18, Punta Cana, Dominican Republic. Association for Computational Linguistics.

\bibitem[{Loukas et~al.(2023)Loukas, Stogiannidis, Malakasiotis, and Vassos}]{loukas2023breaking}
Lefteris Loukas, Ilias Stogiannidis, Prodromos Malakasiotis, and Stavros Vassos. 2023.
\newblock Breaking the bank with chatgpt: Few-shot text classification for finance.
\newblock \emph{arXiv preprint arXiv:2308.14634}.

\bibitem[{Mai et~al.(2019)Mai, Tian, Lee, and Ma}]{mai2019deep}
Feng Mai, Shaonan Tian, Chihoon Lee, and Ling Ma. 2019.
\newblock Deep learning models for bankruptcy prediction using textual disclosures.
\newblock \emph{European journal of operational research}, 274(2):743--758.

\bibitem[{Manning et~al.(2008)Manning, Raghavan, and Sch\"{u}tze}]{manning}
Christopher~D. Manning, Prabhakar Raghavan, and Hinrich Sch\"{u}tze. 2008.
\newblock \emph{Introduction to Information Retrieval}.
\newblock Cambridge University Press, USA.

\bibitem[{Mayew et~al.(2015)Mayew, Sethuraman, and Venkatachalam}]{mayew2015md}
William~J Mayew, Mani Sethuraman, and Mohan Venkatachalam. 2015.
\newblock Md\&a disclosure and the firm's ability to continue as a going concern.
\newblock \emph{The Accounting Review}, 90(4):1621--1651.

\bibitem[{Odom and Sharda(1990)}]{odom1990neural}
Marcus~D Odom and Ramesh Sharda. 1990.
\newblock A neural network model for bankruptcy prediction.
\newblock In \emph{1990 IJCNN International Joint Conference on neural networks}, pages 163--168. IEEE.

\bibitem[{Ohlson(1980)}]{ohlson1980financial}
James~A Ohlson. 1980.
\newblock Financial ratios and the probabilistic prediction of bankruptcy.
\newblock \emph{Journal of accounting research}, pages 109--131.

\bibitem[{Ouyang et~al.(2022)Ouyang, Wu, Jiang, Almeida, Wainwright, Mishkin, Zhang, Agarwal, Slama, Ray et~al.}]{ouyang2022training}
Long Ouyang, Jeffrey Wu, Xu~Jiang, Diogo Almeida, Carroll Wainwright, Pamela Mishkin, Chong Zhang, Sandhini Agarwal, Katarina Slama, Alex Ray, et~al. 2022.
\newblock Training language models to follow instructions with human feedback.
\newblock \emph{Advances in Neural Information Processing Systems}, 35:27730--27744.

\bibitem[{Wolpert(1992)}]{wolpert1992stacked}
David~H Wolpert. 1992.
\newblock Stacked generalization.
\newblock \emph{Neural networks}, 5(2):241--259.

\bibitem[{Zhuang et~al.(2021)Zhuang, Wayne, Ya, and Jun}]{zhuang-etal-2021-robustly}
Liu Zhuang, Lin Wayne, Shi Ya, and Zhao Jun. 2021.
\newblock \href {https://aclanthology.org/2021.ccl-1.108} {A robustly optimized {BERT} pre-training approach with post-training}.
\newblock In \emph{Proceedings of the 20th Chinese National Conference on Computational Linguistics}, pages 1218--1227, Huhhot, China. Chinese Information Processing Society of China.

\end{thebibliography}
\bibliographystyle{support/acl_natbib}

\appendix
\section*{Appendix}
\label{sec:appendix}

\begin{figure}[h!]
\vspace{-0.05cm}
    \centering
    \includegraphics[width=0.40\textwidth]{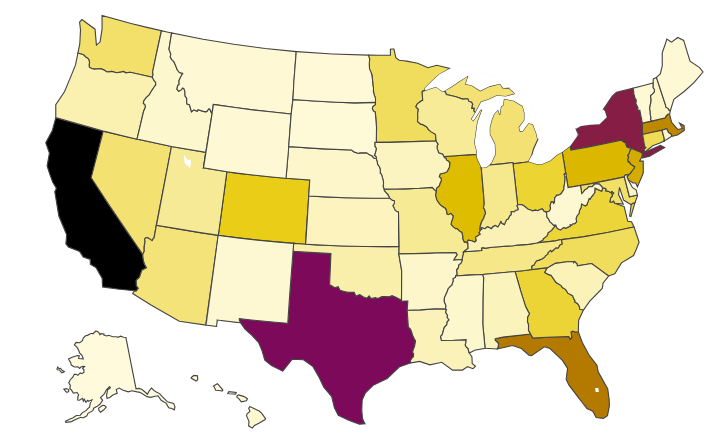}
    \caption{Distribution of the headquarters for the companies in ECL across the United States. A darker colour indicates that more firms are located in this state.
    }
    \label{fig:map}
\end{figure}

\begin{figure*}[]
\centering
    \includegraphics[width=\textwidth]{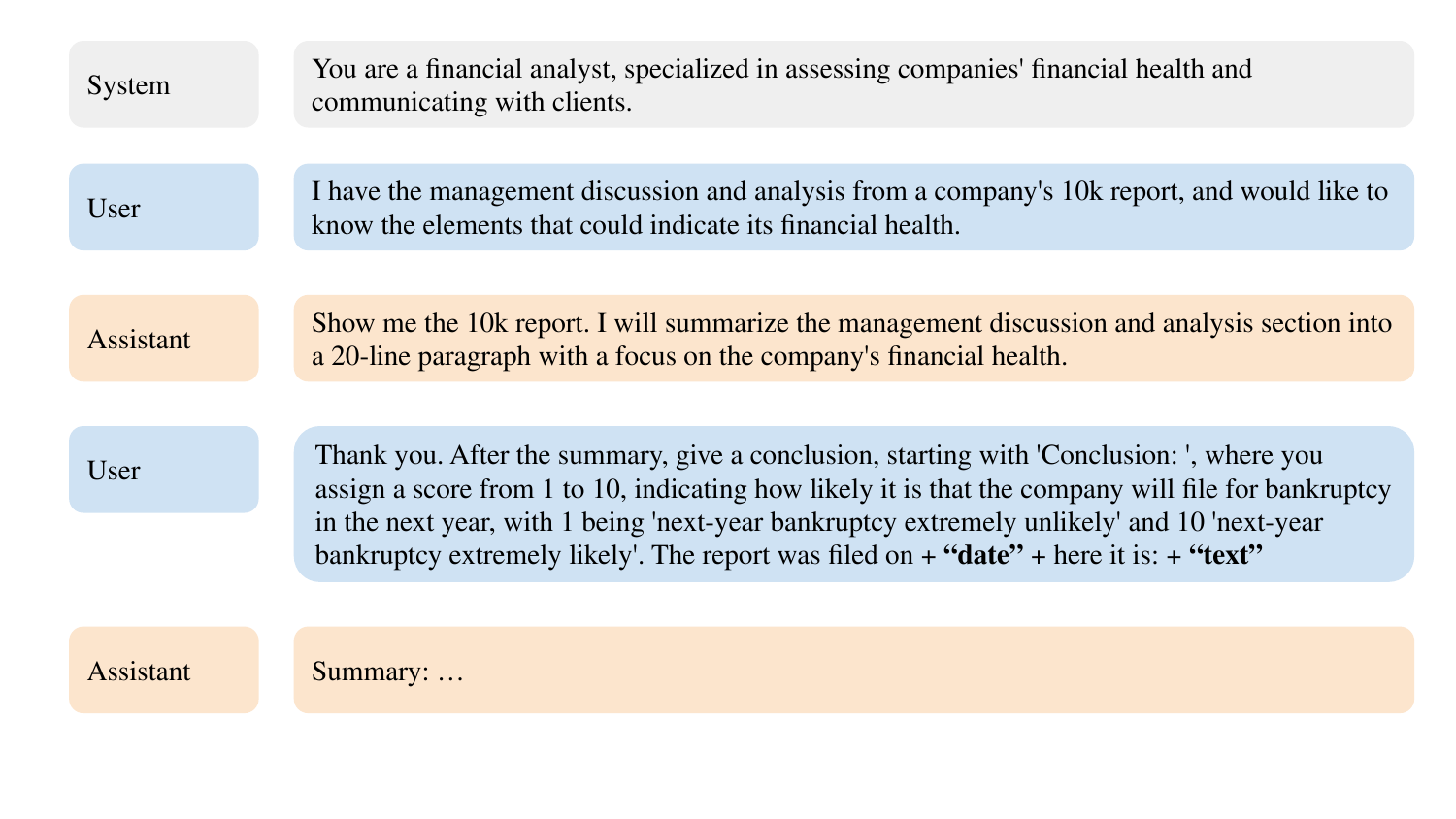}
    \caption{The prompt given to GPT-3.5 Turbo (with a context window of 16,000 tokens) for (1) summarisation of the MD\&A section of the 10K's and (2) zero-shot bankruptcy prediction. }
    \label{fig:prompt}
\end{figure*}

\begin{figure}[]
    \centering
    \includegraphics[width=0.5\textwidth]{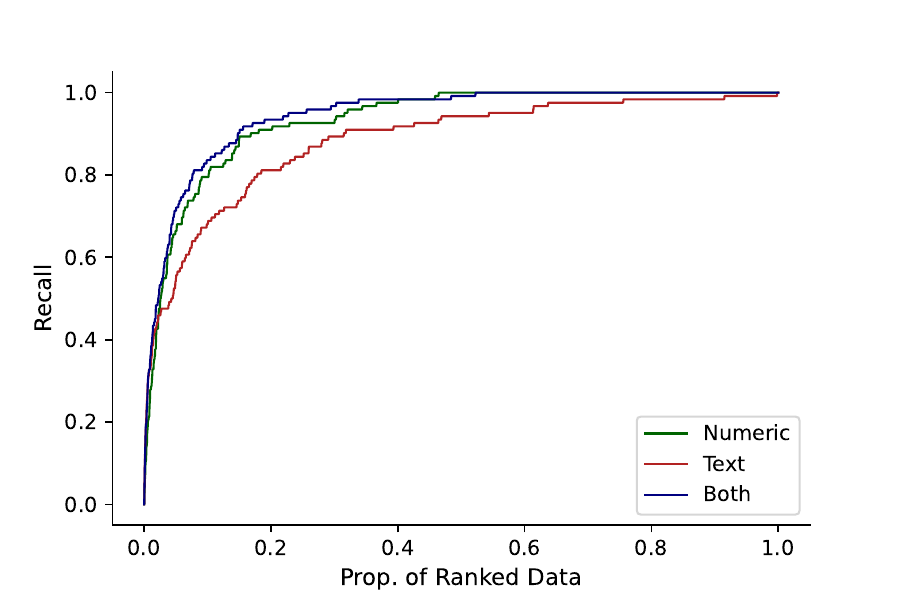}
    \caption{Cumulative Accuracy Profile (CAP Curve) for the best textual (TF-IDF), numerical (XGBoost) and combined (ensemble) models evaluated on the test set. Note that this curve is similar to the ROC curve due to the class imbalance.}
    \label{fig:CAP}
\end{figure}

In the Appendix we show a map with the distribution of the headquarters for the companies in ECL in Figure~\ref{fig:map}. Table~\ref{tbl:sectors} presents the distribution of the industries for the companies in ECL. The prompt given to GPT-3.5 for summarisation and zero-shot bankruptcy prediction is shown in Figure~\ref{fig:prompt}. The CAP curve for the best numerical, textual and combined models is shown in Figure~\ref{fig:CAP}. Table~\ref{tbl:description} contains a description of the numerical variables and Table~\ref{tbl:overviewECL} gives an overview of ECL and the dataset splits.

\begin{table}[]
\footnotesize
\caption{\label{tbl:sectors} Distribution of the industries (SIC divisions) for the companies in ECL. Most companies are active in the manufacturing industry (shown in bold), followed by finance, insurance and real estate.
}
\begin{tabular}{@{}lc@{}}
\toprule
\textbf{SIC Division}              & \textbf{Proportion of Data} \\ \midrule
Agriculture, Forestry, Fishing     & 0.35\%                \\
Mining                             & 4.61\%                \\
Construction                       & 1.02\%                \\
\textbf{Manufacturing}                      & \textbf{37.14}\%               \\
Transportation \& Public Utilities & 9.21\%                \\
Wholesale Trade                    & 3.04\%                \\
Retail Trade                       & 5.09\%                \\
Finance, Insurance, Real Estate    & 20.95\%               \\
Services                           & 17.01\%               \\
Public Administration              & 1.58\%                \\ \bottomrule
\end{tabular}
\end{table}

\begin{table*}[]
\footnotesize
\begin{tabular}{@{}p{5.5cm}p{2cm}p{5.5cm}p{2cm}@{}}
\toprule
\textbf{Variable}                                       & \textbf{CompuStat} & \textbf{Variable}                                   & \textbf{CompuStat} \\ \midrule
Current Assets / Current Liabilities        & ACT / LCT          & Current Liabilities / Sales                   & LCT / SALE         \\
Accounts Payable / Sales                                & AP / SALE          & Total Liabilities / Total Assets                    & LT / AT            \\
Cash and Short Term Investments / Total Assets          & CHE / AT           & Log (Total Assets)                                  & Log (AT)           \\
Cash / Total Assets                                     & CH / AT            & Log (Sales)                                         & Log (SALE)         \\
Cash / Current Liabilities                        & CH / LCT           & Net Income / Total Assets                           & NI / AT            \\
(EBIT + Depreciations and Amortisations) / Total Assets & (EBIT + DP) / AT   & Net Income / Sales                                  & NI / SALE          \\
EBIT / Total Assets                                     & EBIT / AT          & Operating Income After Depreciations / Total Assets & OIADP / AT         \\
EBIT / Sales                                            & EBIT / SALE        & Operating Income After Depreciations / Sales        & OIADP / SALE       \\
{[}Total Debt in Current Liabilities + (0.5)*Total Long Term Debt{]} / Total Assets &
  (DLC + 0.5*DLTT) / AT &
  (Current Assets - Inventory) / Total Current Liabilites &
  (ACT - INVT) / SALE \\
Inventory Decrease / Inventory                          & INVCH / INVT       & Retained Earnings / Total Assets                    & RE / AT            \\
Inventory / Sales                                       & INVT / SALE        & Retained Earnings / Current Liabilities       & RE / LCT           \\
Current Liabilities - Cash / Total Assets         & (LCT - CH) / AT    & Sales / Total Assets                                & SALE / AT          \\
Current Liabilities / Total Assets                & LCT / AT           & Total Equity / Total Assets                         & SEQ /AT            \\
Current Liabilities / Total Liabilities           & LCT / LT           & Working Capital / Total Assets                      & WCAP / AT          \\ \bottomrule
\end{tabular}
\caption{\label{tbl:description} This table presents the numerical variables used by our classifiers and the corresponding formulas in CompuStat. We derived the variables from the work of \citeauthor{mai2019deep} (\citeyear{mai2019deep}) but only include those that can be computed from the 10K and discard the variables that require market information (e.g. stock market returns). 
}
\end{table*}

\newpage


\begin{table*}[]
\centering
\footnotesize
\begin{tabular}{@{}l>{\centering\arraybackslash}p{1.5cm}
>{\centering\arraybackslash}p{1.75cm}
>{\centering\arraybackslash}p{2.2cm}
>{\centering\arraybackslash}p{1.75cm}
>{\centering\arraybackslash}p{1.75cm}
>{\centering\arraybackslash}p{1.75cm}@{}}
\toprule
\textbf{Dataset} &
  \textbf{Number of 10K Filings} &
  \textbf{Period (Filing Year)} &
  \textbf{Average Asset Value (Billion \$)} &
  \textbf{Number of Positives} &
  \textbf{Proportion of Positives} &
  \textbf{Negatives per Positive} \\ \midrule
ECL Complete      & 170,139 & 1993 - 2023 & 1.387 & -   & -      & -   \\
ECL Labelled      & 84,652  & 1993 - 2021 & 3.435 & 662 & 0.78\% & 127 \\
Full Training Set & 66,363  & 1993 - 2015 & 2.851 & 540 & 0.81\% & 122 \\
Training Set      & 54,039  & 1993 - 2011 & 2.518 & 481 & 0.89\% & 112 \\
Validation Set    & 12,324  & 2012 - 2015 & 4.547 & 59  & 0.48\% & 208 \\
Testing Set       & 18,289  & 2016 - 2021 & 5.995 & 122 & 0.67\% & 149 \\ \bottomrule
\end{tabular}
\caption{\label{tbl:overviewECL} This table gives an overview of the ECL dataset and the training, validation and test sets that were used for the next year bankruptcy prediction task. A positive sample refers to a 10K record filed in the year before bankruptcy. The average asset value (in billion \$) is not corrected for inflation and computed after removal of the outliers exceeding the 95\% quantile. Note that we do not include statistics on the label distribution for the next year bankruptcy prediction task for the complete ECL dataset. Some samples in this dataset cannot be assigned a label since they do not qualify for inclusion in the LoPucki BRD.}
\end{table*}

\end{document}